Science, Technology, Engineering, and Mathematics Undergraduates' Knowledge and Interest in Quantum Careers: Barriers and Opportunities to Building a Diverse Quantum Workforce

Jessica L. Rosenberg
*George Mason University, Department of Physics and Astronomy*

Nancy Holincheck, Michele Colandene
*George Mason University, School of Education*

Efforts to build the workforce in support of the second quantum revolution are growing, including the creation of education programs that will prepare students for jobs in this area. We surveyed 186 undergraduate students with majors across the STEM disciplines and followed up with group interviews to understand their perspectives. The project was designed to understand what these STEM students know about quantum and quantum career opportunities and their level of interest in pursuing a career related to quantum. We found that most of the students know very little about quantum. Nevertheless, except for students in the life sciences, there was an interest in quantum careers. Across STEM majors, women were less likely to express interest in quantum careers than men, but this difference disappeared when we examined only physical and computer science majors. Of the few students who had knowledge of quantum concepts, most learned about this topic from online media, especially online videos. Some students reported learning about quantum in high school classes, where it was taught as an extension beyond the usual topics of the course. The undergraduate STEM students in our study identified multiple ways they would like to learn more about quantum, including short videos, seminars, courses, certificates, and degree programs.

## I. INTRODUCTION

America's position at the forefront of research and development in science, technology, engineering, and math (STEM) is a critical driver of the economy and prosperity (e.g., [1]). Staying at the forefront will require more students to pursue STEM careers. Research has shown that early interest and experiences in STEM play an important role in pursuing academic studies and careers in these areas [2].

While research and technology are important to the American economy in general, quantum specifically holds enormous potential to drive innovation across the US economy and transform our technology as we enter the second quantum revolution [3]. The first quantum revolution gave us the transistor, which is at the core of our current economy, powering our electronic devices. This new quantum revolution has been driven by the ability to manipulate quantum states in new ways and opens up opportunities in computing, communications, and sensing. Work to develop the quantum materials that are at the heart of these quantum technologies is also important for future progress. Because the development of quantum technologies is critical for national defense and economic development, the National Quantum Initiative was signed into law in December 2018. It was renewed in May 2022 to prioritize the development of a new high-tech workforce [4,5].

The National Quantum Initiative emphasized the development of the quantum workforce to help prepare for a quantum marketplace that is currently worth $614 million in quantum computing alone and is expected to grow at an annual rate of 25% out to 2025 [6]. A recent survey of employers found that, for 21 companies that identify themselves as being part of the quantum industry, quantum jobs span a range of disciplines (physics, engineering, computer science, chemistry, and math) and require a range of degrees from an associate's degree in engineering to PhDs [7]. Hughes et al. [8], building on the work of Fox et al., examined the expected growth in the quantum industry for 57 companies [8]. They found that most of the companies surveyed expected to be hiring in quantum in the next five years and, as with Fox and colleagues, they found that the hiring was expected to be in a range of disciplines and at a range of education

levels, including some positions related to business which expands the range of background and skills needed in this area.

To meet the growing needs of industry, universities are developing new quantum programs in quantum information science (QIS). Historically, quantum has been part of the upper-level (junior/senior undergraduates and graduate students) physics curriculum. As part of the physics curriculum, efforts to improve the understanding of quantum have examined how students understand quantum mechanics concepts (e.g., [9-13]). This work on the development of conceptual understanding will continue to be important as quantum and the teaching of quantum concepts move to earlier stages and other disciplines. Additional research will be needed to understand how a conceptual understanding of quantum is developed in quantum information science which extends beyond the standard physics curriculum. While most of the QIS courses are listed (or cross-listed) in physics, electrical or computer engineering, or computer science [14], many of these courses are transdisciplinary at their core [7].

Some work has been done to examine what the new quantum workforce needs to know to succeed. In a European study, basic principles or phenomena, mathematics, physical background, and applications were identified as important competence areas for developing the quantum workforce [15]. These results were refined in [16] to include (1) theoretical background, including the concepts and phenomena of quantum physics, classical physics, math, and quantum computer science; (2) practical background, including experimental skills, physical/technical realization, engineering/industrialization, and soft/social skills; and (3) applications, including engineering, production of quantum technologies, applications of quantum computing, communications, sensing, and simulation. Gerke [15] and Greinert [16] highlight the breadth of areas that will be part of the second quantum revolution and lead to a wide range of skills that the new workforce will need. Because of the wide range of applications, "quantum" sometimes refers to only quantum computing while at other times it covers a much wider range of technologies including communication, sensing, and materials. In addition, the boundary between old and new quantum technologies is fuzzy with technologies filling the gap between those in the first- and second-generation. The skill sets needed to support these efforts is broadened even more when the technicians who will support efforts in quantum adjacent fields are considered [17].

As of the 2019/2020 and 2020/2021 school years, 74 institutions in the United States offered courses on QIS [14]. Of the institutions offering the QIS courses, 65 (88%) are doctoral-granting institutions [14], representing only 40% of physics Bachelor's degree graduates [18]. Undergraduate students at primarily undergraduate institutions, which include most of the Historically Black Colleges and Universities (HBCUs) and Minority Serving Institutions (MSIs), are much less likely to have access to a QIS course. In 2021, there were three certificates and seven MS programs in the US [19], almost all of which were at research universities with quantum programs. A web search shows that as of the start of 2023, there are at least six new MS programs and seven new certificates. In addition, there are a few university minors and concentrations as well as students pursuing PhDs focusing on quantum across a range of fields.

With the need for workers in the quantum industry growing [7,8], it will be important to focus on recruiting and retaining a diverse quantum workforce. The leaks in the pipeline and the lack of diversity that currently exists in the physical sciences, math, and computer science, which are the likely feeders into quantum, must not be reproduced. These fields have some of the lowest percentages of women and minorities [20]. These low numbers are driven both by smaller numbers entering these disciplines and by greater losses from these disciplines. As these programs grow, creating a diverse workforce will require engaging students beyond those at R1 institutions who are already looking into physics and computer science at institutions with well-established quantum programs. Understanding how students want to learn about quantum and gain access to the information is vital to this endeavor.

In this study, we examined the attitudes of undergraduate students in STEM towards quantum and quantum careers. We survey and talk to students at George Mason University, a large public university that recently became an R1 and has a medium-sized, but growing quantum program. We aim to understand their interest in quantum and where they are learning about this field so that we can effectively plan and develop programs that will welcome and appeal to a broader group of students. We are particularly interested in the thinking of students who are not in upper-level physics courses, where quantum has traditionally been taught. We aim to learn what these students see as the barriers and opportunities to pursuing this field.

The following research questions directed our inquiry:

1. What knowledge and interest do students have in quantum and quantum careers?
2. What barriers and opportunities do undergraduates report with regard to pursuing quantum careers?
3. Where have students gained knowledge about quantum, and what are their suggestions for program development?

## II. METHODS

We used a mixed methods [21] approach to address our research questions. We first used a survey to collect data from a broad range of STEM students (n=186) at George Mason University. To better understand the quantum interest and knowledge of the future workforce, we selected a subset of these students and conducted group interviews. This allowed us to first gather data from a large number of students about their knowledge and interest in quantum and quantum careers and then engage with students interested in quantum to learn more about their specific interests and beliefs.

### A. Participants

A total of 185 undergraduate STEM students from George Mason University participated in this study between February and December 2022. Mason is a large (39,000 students) research-intensive university in the mid-Atlantic region of the U.S. with a diverse student body. Participants were recruited through posters near lecture halls in the science buildings, emails to course instructors, and emails to STEM student groups. Only students who indicated they were STEM majors at Mason were included in this study. Of the 169 students who provided their specific major, 32% were majoring in physical sciences and engineering (physics, engineering, math, chemistry, geography, and geology), 37% were majoring in computer sciences (computer science, cybersecurity, information technology, and data science), and 23% were majoring in life sciences (biology, psychology, forensics, kinesiology, and neuroscience).

Table I shows the demographics of the students who completed the survey and participated in the focus groups. The population that we sampled was 38% female, 48% male, and 4% non-binary (10% of students did not report their gender). The sample was 40% white, 26% Asian, 7% Black, and 12% Hispanic (5% reported two or more races, and 10% did not report their race or ethnicity). By race and ethnicity, 40% of the surveyed students were white, slightly overrepresenting these students since the participants were drawn from a combination of the Mason College of Science (COS), which is 40% white, and the College of Engineering and Computing (CEC) which is 29% white, and slightly underrepresenting most of the other groups. Males were 48% of the survey population, which is somewhat lower than the Spring 2022 percentages for students from the COS (61%) and the CEC (35%).

Our data collection methods included an online survey of undergraduate STEM students followed by focus group interviews with 13 students purposefully selected from the interview participants. The focus group participants, including their pseudonyms and demographic information, are provided in Table II.

### B. Data Collection and Analysis
#### 1. Undergraduate STEM Survey

We designed the survey (see Appendix for survey questions) to assess undergraduate STEM students' knowledge and interest in quantum careers. As Dillman et al. [22] recommended, both forced-response and open-ended questions were included. The survey sections included (1) demographic data, (2) self-assessed knowledge about quantum and quantum careers, (3) interest in quantum and quantum careers, (4) where students learned about quantum and quantum careers, (5) factors that have influenced their thinking about careers, and (6) feelings about whether quantum careers are risky and how important risk is to them in selecting a career path.

While this paper explores what "quantum" means to undergraduate STEM students, we define "quantum careers" as those that come from the development of new quantum technologies across the spectrum, spanning materials, computing, sensing, and communications. We also acknowledge that the growth in quantum adjacent careers - those that support quantum technologies like electronics, cryogenics, and others - are an important part of the discussion of quantum careers – though we do not explicitly discuss those adjacent pathways here.

The questions related to the importance of specific factors for students' perceived career satisfaction and about the people who influenced students' choice of career (group 5) were drawn from the Sustainability and Gender in Engineering (SaGE) survey [23]. Demographic questions about participants' gender, race or ethnicity, and first-generation college student status were asked as optional, open-ended questions to expand inclusivity [24]. These questions were included at the end of the survey to reduce the impact of these questions on students' other survey responses.

The survey was advertised to STEM students around the university through emails from departments and faculty, social media announcements from student STEM groups, in-class announcements, and flyers placed around the university. A gift card raffle was offered to students who completed the survey to encourage participation. The majority of participants responded in the first four weeks of the study, between February and March 2022. Additional students completed the survey in the months that followed; we considered all responses received by December 2022 in our analysis of survey data. Table I shows the demographics of the survey respondents.

Table I. Demographics of survey participants, including number and percent of the total in each category by race, gender, and major.

|  | Total | Male | Female | Non-binary | Gender not reported |
|---|---|---|---|---|---|
| Total | 185 | 88(48%) | 71(38%) | 7(4%) | 19(10%) |
| White | 73(39%) | 42(58%) | 25(34%) | 4(5%) | 2(3%) |
| Asian | 48(26%) | 26(54%) | 21(44%) | 1(2%) | 0(0%) |
| Black/ African American | 13(7%) | 3(23%) | 10(77%) | 0(0%) | 0(0%) |
| Hispanic/ Latino | 22(12%) | 10(45%) | 10(45%) | 2(9%) | 0(0%) |
| Native American | 1(1%) | 0(0%) | 1(100%) | 0(0%) | 0(0%) |
| Multi- racial | 9(5%) | 5(56%) | 4(44%) | 0(0%) | 0(0%) |
| Race/ ethnicity not reported | 19(10%) | 2(11%) | 0(0%) | 0(0%) | 17(89%) |
| Physics, Eng, Chem, Math, Geo | 59(32%) | 34(58%) | 22(37%) | 2(3%) | 1(2%) |
| Cyber, CS, IT Data Sci | 68(37%) | 44(65%) | 21(31%) | 2(3%) | 1(2%) |
| Bio, Psych, Forensics, Kinesiology, Neurosci. | 42(23%) | 10(24%) | 28(67%) | 3(7%) | 1(2%) |

*2. Focus Group Interviews*

The research team developed the semi-structured focus group interview protocol to help us learn more about student knowledge and interest in quantum and quantum careers, how students learned about quantum and quantum careers, student beliefs about quantum and quantum careers, and the student's views on the best ways to learn about these topics.

In March of 2022, a subgroup of survey participants was identified for focus group interviews. Students were selected for interviews based on their responses to specific survey questions, including "I am interested in pursuing a career related to quantum." We focused on students who expressed agreement with this statement, which we interpreted as responses of 3 or above on a continuous sliding scale ranging from 1 (strongly disagree) to 5 (strongly agree). From the group of students who expressed interest in a quantum career, we pseudo-randomly selected students with a range of undergraduate STEM majors, including students from across the physical sciences, computer sciences, and life sciences. After identifying twenty students who met the selection criteria, we invited them by email to sign up for one of the group interviews, which were offered at a variety of times. Several students were unavailable at the times interviews were scheduled, and some students signed up but did not attend. We conducted four Zoom focus group interviews with 13 students. All students who participated in the focus group interviews received a $25 gift card. Interviews were recorded and transcribed for analysis. Table II provides a pseudonym for each of the focus group participants, along with demographic information. Each focus group included students from multiple STEM majors and years in school.

Table II: Focus group participant information.

| Pseudonym | Gender | Race | Year | Major | First Gen | Focus Group | Interest in pursuing a quantum career | Knowledge of quantum careers |
|---|---|---|---|---|---|---|---|---|
| Samir | Male | South Asian | Soph. | Comp. Sci. | Yes | 1 | 5 | 2 |
| Benjamin | Male | White/mixed | Senior | Bioeng., Math | No | 1 | 5 | 3 |
| Heather | Female | White | Senior | Comp. Sci. | No | 1 | 4.1 | 1 |
| Shriyan | Male | South Asian | Soph. | Physics | Yes | 2 | 4.1 | 0 |
| William | Male | White | 5th year | Cyber Eng., Math | Yes | 2 | 4 | 2 |
| Wyatt | Male | Half African American | Soph/Junior | Comp. Sci. | No | 2 | 3.6 | 1 |
| Vicente | Male | Hispanic | Junior | Comp. Sci. | Yes | 3 | 3 | 0 |
| Gail | Female | White | Soph. | Forensic Sci. | No | 3 | 3 | 1 |
| Harry | Male | White | Fresh. | Comp. Sci. | No | 3 | 3.5 | 2 |
| Jabar | Male | Asian - Indian | Junior | Comp. Sci. | No | 3 | 3 | 0 |
| Faariq | Male | Middle Eastern | Junior | Electrical Eng. | No | 4 | 3 | 0 |
| Reva | Female | Asian | Soph. | Comp. Sci. | No | 4 | 3.5 | 0 |
| Abdul | Male | Asian | Soph. | Comp. Sci. | Yes | 4 | 3.8 | 1 |

*3. Data Analysis*

Survey data were cleaned and coded for analysis in SPSS (version 28). Descriptive statistics and crosstabs analyses were used in the initial analysis of survey data, followed by content analysis [25] to code responses to the open-ended questions on the survey. Independent samples t-test and Analysis of Variance (ANOVA) were used to examine differences in continuous variables. Effect sizes were calculated to interpret the magnitude of statistically significant differences.

Group interviews were transcribed for analysis and checked for accuracy. We engaged in collaborative coding of student interviews using both descriptive and in-vivo coding methods [26]. Each interview was coded independently by both Holincheck and Rosenberg and then discussed during a research meeting. After all interviews were coded, we compiled our codes to use focused coding to consolidate codes and identify themes. Our findings for each research question are discussed below.

III. FINDINGS

As discussed below, our findings offer insight into what STEM undergraduate students know about quantum and quantum careers, their interests in quantum careers, and how they have previously learned about quantum. A large majority of students knew little or nothing about quantum, yet more than half expressed some interest in pursuing a career in quantum. Students from underrepresented groups (by gender and race/ethnicity) were less likely to express an interest in quantum than those traditionally overrepresented in STEM majors and careers. Students who reported learning about quantum reported doing so from various sources, including online media, books, and high school or college classes.

Our survey of students included general questions about knowledge of and interest in quantum careers as well as questions that asked more specifically about quantum computing and its applications, quantum cryptography, quantum sensing, and quantum materials. Because students knew so little about quantum overall, we focused on the general questions about quantum careers.

A. Undergraduate Students' Knowledge and Interest in Quantum and Quantum Careers

To understand what knowledge and interest students have in quantum and quantum careers (research question 1), we analyzed both the survey and focus group interview data. The pie chart in Figure 1 shows the distribution of student responses to the question, "What do you know about quantum careers in general." Most students (88%) said that they knew "Nothing" (55%) or "A little bit" (33%). The knowledge of quantum careers did not differ significantly when subdivided by major, gender, race, or first-generation college student status.

Despite their lack of knowledge, the histogram in Figure 1 shows that students were moderately "interested in pursuing a career related to quantum," with 63% rating their interest a 3.0 or higher on a continuous 5-point scale (five being "strongly agree" and one being "strongly disagree"). However, unlike for knowledge, there were statistically significant differences in interest by gender (see Figure 2) and race/ethnicity but no significant difference due to first-generation status. Note that Figure 2 shows histograms of the responses by gender and discipline (lower plots) as well as the response values for each person (upper plots).

We conducted an independent samples t-test on the continuous variable (interest in pursuing a career related to quantum) to determine how gender played a role in students' interest in quantum careers. We found a statistically significant difference based on gender. Across all majors, students who identified as men (n = 68, M = 3.256, SD = 1.01) reported a greater interest in pursuing a career in quantum than students who identified as women (n = 49, M = 2.608, SD = 1.17), t(115) = 3.201, p=0.002. The mean response for the five students who identified as non-binary (M = 2.860, SD = 1.57) was not statistically significantly different from either men or women. Because non-binary students and women are both underrepresented in STEM [27], we conducted a t-test to compare students who identified as men with students who identified with genders underrepresented in STEM (women and non-binary students). Across all majors, students who identified as men (n = 68, M = 3.256, SD = 1.01) reported a greater interest in pursuing a career in quantum than students who identified with genders underrepresented in STEM (n = 54, M = 2.63, SD = 1.20), t(103.48) = 3.063, p=0.003. This indicates that men indicated a higher interest in quantum careers than other gender groups.

The life science majors in our study were mostly women, so we examined whether the difference in interest based on gender was related to their major rather than gender alone. To understand the influence of students' chosen major on our results, we created three bins: life sciences (which had a much larger fraction of women than the other two bins, 67%), cyber and computer science, and computer engineering (31% female respondents, we refer to this group as the computer science majors), and physical science, engineering, and math (37% female respondents, we refer to this group as the physical science majors). When the independent samples t-test was repeated with only computer science and physical science majors, we found no statistically significant difference between women (n = 32, M = 3.000, SD = 0.20) and men (n = 62, M = 3.30, SD = 0.13) in their interest in a quantum career, t(92) = 1.289, p = 0.100. This indicates that their major plays a more important role in students' interest in quantum than gender, as women in physical science and computer-related fields have a similar interest in quantum to the men in the study.

To determine the role of race/ethnicity in student interest in a quantum career, we conducted an independent samples t-test and found a statistically significant difference based on race/ethnicity. Across all majors, students with a racial or ethnic background that is historically marginalized in STEM (Black, Hispanic, and Native American; n = 30, M = 2.433, SD = 1.00) had less interest in pursuing a career in quantum than students who were not from a racial or ethnic group historically marginalized in STEM (white and Asian; n = 83, M = 3.07, SD = 1.13, t(111) = -2.693, p=0.008). To investigate whether this was related to the differences we found for life science majors, we repeated this analysis for only computer science and physical science majors and again found a statistically significant difference based on race/ethnicity. For computer science and physical science majors, students from a group that is historically marginalized in STEM (Black, Hispanic, and Native American; n = 22, M = 2.532, SD = 1.09) had less interest in pursuing a career in quantum than students who were not from a racial or ethnic group historically marginalized in STEM (white and Asian; n = 67, M = 3.35, SD = 0.99, t(87) = -3.269, p<0.001). In contrast to our findings for gender, accounting for student majors did not remove the difference in interest in quantum careers based on race/ethnicity.

To determine differences in interest in quantum based on first-generation college student status, we again used an independent samples t-test. We found no statistically significant difference in interest in pursuing a career in quantum for students who were first-generation college students (n = 44, M = 2.971, SD = 1.18) as compared to students who are not (n = 80, M = 2.975, SD = 1.12, t(122) = -0.021, p=0.983).

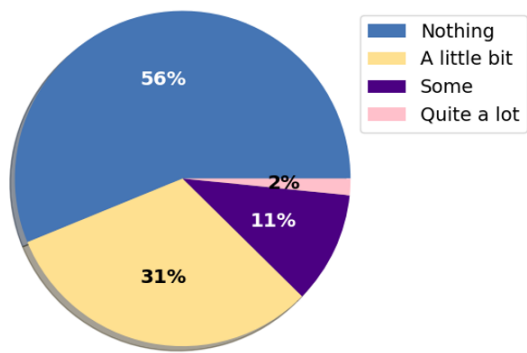
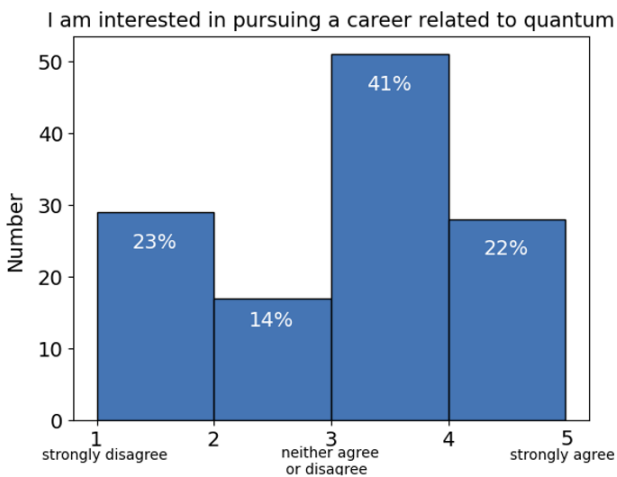

FIG. 1. The pie chart shows students' self-described knowledge of quantum careers. The histogram shows the number and percentage of students in each bin with respect to how they self-described their interest in pursuing a career related to quantum with students rating their interest on a continuous scale between 1 (strongly disagree) and 5 (strongly agree). The four bins are (1-1.9), (2-2.9), (3-3.9), (4-5). The y-axis is the number of students, and the bin labels show the percent with respect to the total number (125) of students who answered the question.

In the survey, we asked students several open-ended questions, including "Briefly summarize what you know about careers in quantum." The largest number of students (69 students or 42%) reported knowing nothing about quantum, while 11% (18 students) did not respond, and 10% (16 students) responded by referencing STEM topics without discussing quantum careers (i.e., robotics, quantum superposition). The other 37% (61 students) identified current or future quantum careers or discussed their general knowledge about quantum careers. We coded these responses and sorted them thematically.

Of the 61 students who displayed knowledge of quantum careers, 44% (27 students) described quantum careers as related to computing. A geography major stated, "They involve computing that is beyond simply binary and are as such far more powerful." A mathematics major described quantum careers as "Mostly just things in quantum computing but nothing more."  A computer science major demonstrated his understanding with the response, "There is a race to maximize stable qubits in quantum computing." Most of these students described the work that would be done with quantum computers rather than specific jobs or careers.

A nearly equal number of students (26 or 43%) described quantum careers as related to theory and research, often referencing university research. One computer science major replied, "I know that most careers having to do with quantum are research-oriented and university-affiliated." An electrical engineering major stated, "I know that careers in quantum are very difficult to understand, and most are research-oriented in order to learn the basics about the properties of how our matter and energy works."  Similarly, a forensic science major responded, "The field of quantum strives to understand the nature of the universe… or at least I think they do.  I know that to be a quantum physicist, it takes a PhD and even postdoctoral work." This group of students described quantum jobs as university-based and theory-oriented rather than referencing current industry jobs in quantum.

Only a small number of students who displayed knowledge of quantum careers (8 students, or 13%) captured ideas about careers that included references to industry careers (including those within computing) and research. A computational physics major offered his response, "There are careers in quantum available in both academia and industry. On the academia side, this involves teaching college students about quantum physics as well as performing academic research into various fields of quantum physics. On the industry side, this involves performing research into quantum physics as well as direct applications of quantum theory." Similarly, a mathematics major wrote, "There are theoretical physicists, those who are working on combining the view of quantum with other views and concepts such as quantum gravity. There's also a big push towards the more application-based idea of creating a quantum computer, which needs engineers, physicists, and computer scientists."

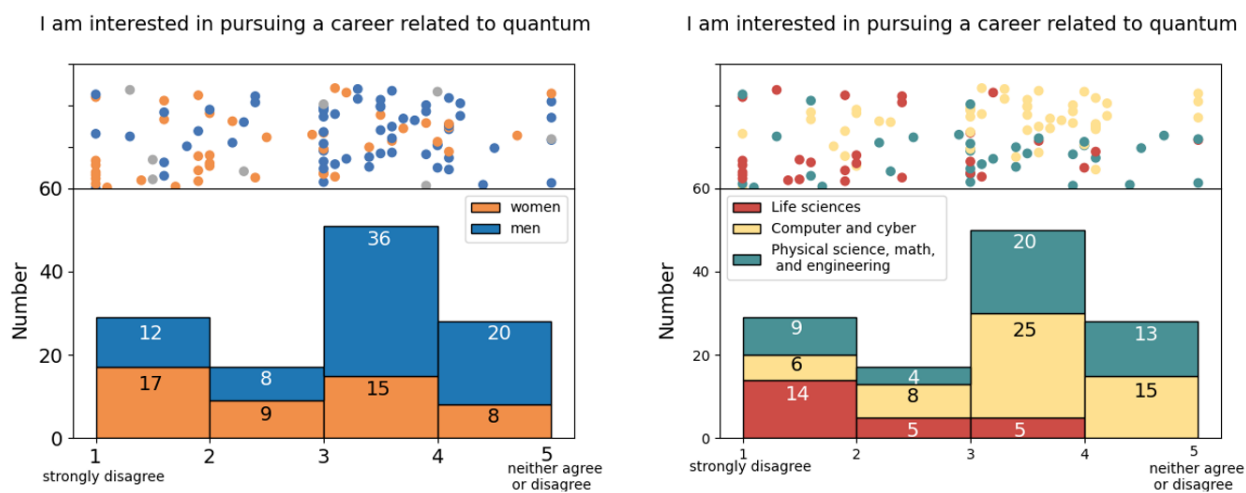

FIG. 2. Histograms showing student survey responses to "I am interested in a career related to quantum" divided up by gender (left) and by students' major (right). The four bins are (1-1.9), (2-2.9), (3-3.9), (4-5). The numbers on the bar chart indicate the number of students in each bin. The scatter plots show the full data sample distribution. Note that the y-axis in the upper plot is arbitrary - points were assigned y-values randomly to make them visible.

In advance of our analysis of focus group interview data, we examined the survey responses for the 13 students who participated in the focus groups. Five (out of 13, 38%) of the focus group students knew "nothing" about quantum careers, four (31%) knew "a little bit," three (23%) knew "some" and one (8%) knew "quite a lot." Despite claiming to know very little about quantum careers, all the focus group students rated their interest "in pursuing a career related to quantum" to be at least a three out of five (our criterion for selecting them for the interviews), with 2 of the students rating their interest a 5, and three rating it 4 or 4.1 on the numerical slider (minimum of 1.0 and maximum of 5.0). Eight of their survey responses included fairly coherent descriptions of what they "know about quantum careers." One of the most comprehensive was given by Heather, who claimed only to know "a little bit." She said, "There is work related to development and applications of quantum computing for both hardware and software. There is also the field of quantum communications, which has large research areas related to space operations."

Despite many of the focus group students providing fairly coherent descriptions of quantum careers in the survey, during the focus group discussions only a few participants could put together a solid description of what quantum is and, after a bit of discussion, describe what quantum careers might entail. It is not clear what drove this difference, but it might be related to the ability to look up information about quantum careers on the survey. Vincente commented, "I've only ever heard the term [quantum] in TV shows and movies and stuff, and I get the idea that it's something like very small as Harry just described very tiny particles or something, maybe even theoretically I'm not sure that's kind of what I think of" and noted that those TV shows and movies "make it sound so hysterically complicated and overly theoretical." Even among these students who were the most interested in quantum, there was a lot of uncertainty as to what it entails. Several mentioned ideas to the effect that they "hardly know much about quantum careers but am interested in learning."

The lack of a deep understanding of the nature of quantum and quantum careers by focus group participants reflects the broader theme seen in the survey responses, like the student who said, "I only know a little bit, and physics definition of quantum, but I want to learn more about it."

### B. Barriers and Opportunities to Students Pursuing Quantum
#### 1. *Factors students selected as influencing their career interests*

In addition to questions specifically related to quantum, our survey asked students to identify people who had influenced their career path and the importance of specific factors for their future career satisfaction. We focused our analysis on the 75 students who indicated they were interested in a quantum career (i.e., responding to this slider question with a rating of 3.0 or higher). This allowed us to consider who influences undergraduate students to consider a quantum career, and the factors students place importance on when choosing a career.

Students were asked to select who was influential in their career interests out of a list of 15 options. The people most often selected as contributing to students' interest in their future careers included a parent or guardian, a science teacher or professor, and someone who works in the industry. Teachers and professors in other disciplines were also influential, including those in computer science, engineering, and mathematics. Only five students indicated that no person had influenced their choice of career.

Questions about factors important to future career satisfaction were drawn from the Sustainability and Gender in Engineering (SaGE) survey [23], and provided students with a list of factors from which to choose. When responding to what factors they believed would be important for their future career satisfaction, they placed importance on developing new knowledge and careers, having job security and opportunity, making use of their talents and abilities, and helping others. They were less likely to place importance on becoming well known or having an easy job.

#### 2. *Challenges discussed by focus group participants*

One of the challenges that must be overcome if a broader group of students is going to pursue quantum is the idea that quantum is only accessible to "geniuses." Leslie et al. [28] showed that the fields with the smallest numbers of black and female PhDs tend to have a greater emphasis on the importance of brilliance for success. Math, physics, computer science, and engineering all rate high in this emphasis on brilliance and have small percentages of black and female PhDs [28].

The impact of this emphasis on genius on students in quantum is evident in some of the comments made by the focus group participants like Reva, who said, "I guess right now, for me, when I think of quantum computing and all that stuff, it's just a mystery to me. It's just something that super geniuses talk about in their free time, doing it as a career. I definitely feel like they have to have a really good understanding of what quantum physics is and then how that applies to computing and tech and all that stuff, and it's probably also very math-heavy." In contrast, Harry noted that it is not as difficult as its reputation, "I think quantum actually has a more scary reputation than it deserves because, honestly, the math isn't as impossible as people make it out to be in the TV shows and movies. I mean it's obviously very difficult, but I don't think it's this super scary thing because, once you dive into it, a lot of it is very intuitive and very beautiful as well as also being scary. So I think it's definitely something worth pursuing further if you're interested in it."

Students also had the sense that the field requires advanced education, "You probably need a Ph.D. to realistically get a career in the field, and even then, the positions will be competitive." and "There's not a wide range of jobs in the field …you're taking a risk studying it.." and from Faariq, "I don't think it could be as practical as math or physics."

Other barriers that students mentioned included the lack of women in the field. Reva noted, "Just getting more women in tech, I feel like, it's the first biggest hill before even broaching quantum computing and stuff like that. I guess to me it's like where to even start introducing quantum computing when a lot of women don't even know or get into computer sciences."

Some of the opportunities in quantum were also evident from the interviews, including the possibility that there would be more quantum jobs in the future. Vicente noted, "Currently, there's not a wide range of jobs in the field. It's always gonna be a risk you're taking, the risk studying it with the hopes that it will [...] be part of the future, so it is risky in that way, the way that we're advancing technologically, it's not the biggest risk in the world."

Beyond the job opportunities, quantum has an opportunity to capitalize on the fact that it is cool and interesting. Jabar notes, "It's also kind of a realm where normal physics kind of just gets thrown out the window. And that was something that kind of stuck with me, and I was like, wow that's really interesting because there's so much going on in the world, and some of it we just can't even explain. As much as we can explain, it's really interesting to me, and I'm constantly interested in learning more about it as well."

### C. Where Students Have Learned about Quantum

To understand where students learned what little they know about quantum (research question 2), we asked our focus group participants. Most of the students surveyed have not learned about quantum careers in either formal or informal settings. For the minority who have some knowledge of quantum and the associated careers, most learned about the subject online, with a few learning about it in high school or college courses. The focus group participants were more likely to have learned about quantum than the overall survey population. Several interviewees learned about quantum in college courses, including physics, chemistry, and cybersecurity, as well as in a high school class and, for Abdul, in the online Qubit x Qubit course. In discussing his physics class, Jabar appreciated that the teacher added a discussion of quantum, "I appreciated that my teacher in high school that actually took the time out to show something related to anything quantum… it wasn't even part of the curriculum or anything, it was purely just extra we had an extra day and he thought that it was something interesting and he wanted to show us. I may not have understood everything, but it at least set up something in my mind that I still think about to this day, and that was what five or six years ago, maybe. But it leaves an impact when you just have something presented without any ulterior kind of this is not for a grade or something just pure interest."

The focus group participants were asked where they would like to learn about quantum and there was no general agreement on whether small doses of quantum (workshops, seminars, etc.) or larger doses like full courses, majors, and minors are the answer. Some of the students were looking for smaller doses to fit into busy schedules, while others wanted full courses to understand the topic. It is likely that both will need to be part of the answer to create a quantum-ready workforce.

Abdul, who participated in the Qubit x Qubit course, mentioned the need for mentors, "for me, it was mentors, people who were doing stuff like this place if we didn't have people who are doing quantum computing, I don't have any to look up to so people tell me about it. So, if no one told me oh, there's such a thing as quantum computing, I would never have done it." If we want to draw students into quantum from institutions that are not research intensive, which tend to include HBCUs and MSIs [29], efforts will need to be made to ensure students across STEM disciplines know what quantum is and provide accessible pathways into the field.

The focus group participants also called attention to the importance of introducing students to quantum at a younger age through interactive activities, "I feel like getting access to… the coding programming Scratch, what MIT made that… elementary school started using them to get children more interested in, for example, I didn't know what coding was like until seventh or eighth-grade math but my brother he got introduced to code from Minecraft." As Gail noted, "I think the exposure is the most important. We were talking about how quantum seems like a very scary subject to look into. It's, well, it's out of my hands because I don't understand physics at all, so I think exposure in terms of [at a] younger age, introducing concepts of quantum that can slowly get more detailed and maybe specific. But an application for real-world problems in the simplest sense, I think, will be beneficial because quantum can be applied to a lot of things."

## IV. DISCUSSION, IMPLICATIONS, AND RECOMMENDATIONS

### A. Learning about Quantum Information Science

The students we talked to pointed to the need for a variety of access points to learn about quantum. Some of them were interested in short virtual learning (they were just coming out of COVID and tend to get a lot of their information from YouTube and other online media), while others were looking for courses and certificates or minors. With differences in knowledge, interest, and background, there is no one-size-fits-all answer to what quantum education should look like. We also need to acknowledge that, at least in the near future, most undergraduate institutions will not be able to create courses that introduce students to quantum and quantum careers beyond the physics curriculum.

Regardless of the duration and types of engagements, good and engaging professors will be important to bring in and retain students who begin with a bit of fear of the complexity. Introduction at the high school level is a way to get students engaged, but the highly structured curriculum in the US and many other countries makes it a challenge for teachers to find easy places to add it to their curriculum [30]. Meyer et al. [31] found that instructors for QIS courses at the university level approached the topic in a variety of different ways even when the course emerged from the same discipline (e.g., CS) at the same level (e.g., beyond first-year undergraduate courses).

Several students expressed the belief that working in quantum requires a Ph.D. The challenge with this belief is that, for the moment, many jobs do require PhDs, yet it is expected that the demand for a wider range of educational backgrounds will grow. Because of the uncertainties in what careers are available and will be available in the near future, students need guidance about what they can do with a quantum education and how to get there, and this guidance will have to change with the changing job market.

Students in our study also talked about quantum being for "geniuses," which is particularly problematic as prior research in STEM indicates that fields with this reputation tend to award the smallest numbers of Ph.D.s to black and female students [28]. If we are to create pathways into quantum for a broader array of students, we must find inroads beyond physics classes and find ways to open up opportunities beyond our research university campuses.

### B. Demographic differences

Our sample of students had majors across the STEM disciplines. Overall, the students in the life sciences were less interested than others in quantum. This lack of interest by life sciences students poses a challenge and an opportunity. The lack of interest is a challenge for those who are working at the intersection of biology and quantum particularly with work in quantum sensors (e.g., [32]) and future prospects in quantum computing including with respect to drug discovery (e.g., [33]). The higher percentage of women in these fields is also an opportunity if we can interest these students in quantum.

Women and students from minoritized groups were less likely to be interested in quantum than white and Asian men and are generally underrepresented in the physical and computer science fields that are the natural feeders into quantum. Because of this, focus will need to be placed on how to engage and encourage their interest and connect quantum education to applications and ways quantum can be used to solve grand challenges (and help others).

C. Limitations

Our sample of students represent a wide range of STEM disciplines at a large, public, R1 institution with a diverse student population and a young but growing quantum program. Nevertheless, the results are from a single institution so the results may not be applicable in all settings. Although the sample is not random, the demographics of the participants were similar to those of STEM majors at this university. In our quantitative analysis of survey responses we sorted students by major into life sciences, physical sciences, and computer sciences. This allowed for meaningful analysis of students by disciplinary area because our sample size did not allow us to compare students by individual majors. However, the unique nature of specific STEM majors were not able to be considered in this study, and the characteristics of the majors may overlap with more than one of the groupings blending the differences and distinctions between them.

D. Recommendations

There is a lot of discussion about developing quantum technology for the public good. As Roberson et al. [34] point out, more will have to go into these efforts than just the competitiveness and national security needs for a broad swath of the population to benefit. In the discussion below we identify recommendations for next steps based on the findings of our study. We try to consider, in this discussion that is based on what we have heard from one set of students at one university, how we can make sure there is broad access to this field and that the developments that come from it are not only focused on the more immediate national security issues but also on the human grand challenges that it has the potential to impact. We offer these recommendations for the development of quantum courses and curricular materials.

1. *Connect quantum to ways advances in this field can help others.* Many of the students that we surveyed and interviewed find the idea of quantum and careers in this field interesting even if they don't really understand what that means which provides an opportunity to draw students in. In our study, students identified helping others as one of the key factors they will consider when selecting a career. This finding is consistent with the work of Verdín et al. [23] that showed that among engineering students, women had an above-average interest in helping people. Quantum educators need to emphasize the areas in which quantum can impact the public and provide students with an understanding of the ways to use quantum to help others.

2. *Identify ways to introduce students to quantum topics that are related to their fields of interest.* Our findings indicate that students who are interested in life sciences do not think that they would be interested in quantum. Nevertheless, there are connections between their field of interest and efforts underway in developing quantum sensors for medical uses and the prospect of quantum computing for use in drug discovery among other things (e.g., [32, 33]). These use cases can also be tied into the desire for students to find ways to help others and will be important in making sure that quantum is developed for the public good [34].

3. *Build an understanding of quantum and quantum careers as early as possible.* The students we interviewed talked about the importance of learning about quantum early. There is a need to build an understanding of quantum and quantum careers as early as possible if the field is to be accessible to the widest group of students. This builds on the literature which indicates that improving the demographics in quantum, (and in many of the sciences) is going to require earlier engagement of students from underrepresented groups so that they gain and maintain an interest (e.g., [35, 36]).

4. *"Demystify" quantum, emphasize how interesting it is, and challenge the belief that you have to be a genius to understand it.* Quantum is a difficult topic that relies on a deep understanding of mathematics and physics for a full understanding. Students talked about quantum as being something that geniuses do, which poses a challenge in recruiting them into the field. However, a lot of work is being done to identify ways to teach quantum concepts without all of the math and physics (e.g., [37, 38]) and to identify the needed background for quantum computing and other aspects of this new discipline (e.g., [39]).

5. *Develop information and pathways into quantum that are accessible to students.* One of the challenges that students mentioned was learning about quantum in the ways that they were most comfortable with and best fit into their schedule. The ways students want to learn about the subject differ, so they need information and accessible pathways that meet them where they are, ranging from videos to courses to majors, in order to make quantum accessible at all 2- and 4-year institutions. This is particularly important for quantum because most of the existing courses are being taught at Ph.D. granting institutions and thus are not accessible to the majority of undergraduate students in the U.S.

6. *Develop mentorship, advising, and cohort-building opportunities to keep students engaged and on the right track.* Students in our study described uncertainty about how to pursue a quantum education and/or career. Some also saw the pursuit of such a new field as risky, and they placed a high value on job security. Building the quantum workforce will require support for students as they navigate these issues in a field that is rapidly growing and changing. Students also point to the influence of parents, teachers, and industry professionals in identifying their career path so identifying ways to help students (and their parents) engage with experts in the field can discuss the options that are available and may help expand the pipeline.

There is significant work that remains to be done to understand the views of quantum among students beyond this initial sample. Future work should investigate how these results might differ with different kinds of institutions and more particularly how specific interventions can alter student views of quantum and quantum careers.

# APPENDIX: SURVEY QUESTIONS

*Mark oval 1 (nothing), 2 (a little bit), 3 (some), or 4(Quite a lot):*
- What do you know about quantum careers in general?
- What do you know about quantum careers within your discipline(s)?

*Type in the box:*
- In what classes have you learned about quantum applications and/or career options?
- In what classes have you learned about quantum applications and/or career options?
- Briefly summarize what you know about careers in quantum
- What career do you intend to pursue?

*Select all that apply:*
Which of the following people have contributed to your selection of a career path?[1]
- Mother/female guardian
- Father/male guardian
- Sibling(s)
- Other relative
- Coach
- Contact with someone in that major
- Academic advisor
- Math teacher/professor
- Science teacher/professor
- Engineering teacher/professor
- Computer Science teacher/professor
- Other teacher/professor
- Internship supervisor/mentor
- Someone who works in this industry/career
- No one has contributed to my selection of a career path

*Mark oval for each of the bulleted items below: Not at all important, unimportant, neither important or unimportant, important, very important*
How important are the following factors for your future career satisfaction?
- Making money
- Becoming well known
- Helping others
- Supervising others
- Having job security and opportunity money
- Working with people
- Inventing/designing things
- Developing new knowledge and skills
- Having lots of personal and family time
- Having an easy job
- Being in an exciting environment
- Solving societal problems
- Making use of my talents and abilities
- Doing hands-on work
- Applying math and science
- Participating in a cutting-edge field

*Slider response ranging from Not risky (0), Slightly risky (1), Risky (2), Very risky (3)*

- To what extent do you view the pursuit of a quantum career as risky?

---

[1] Items from Sustainability and Gender in Engineering (SaGE) survey [20]

- To what extent does risk impact your selection of a career path?

*Slider response (continuous) ranging from from Strongly disagree (1), Somewhat disagree (2), Neither agree nor disagree (3), Strongly agree (4), Strongly agree (5)*

- I am interested in pursuing a career related to quantum
- I am interested in pursuing a career specifically related to quantum computing and it's applications
- I am interested in pursuing a career related to quantum cryptography
- I am interested in pursuing a career related to quantum sensing
- I am interested in pursuing a career related to quantum materials
- I understand what it takes to pursue a career related to quantum
- I understand what the subject of quantum computing involves
- I understand what the subject of quantum sensing involves
- I understand what the subject of quantum materials involves

*Type in the box*
Is there anything else you would like to share about responses to this survey or about what you know about quantum and quantum careers?